# Two Detector Reactor Neutrino Oscillation Experiment Kr2Det at Krasnoyarsk. Status Report.

Yu. Kozlov, L. Mikaelyan, V. Sinev

*Russian Research Center Kurchatov Institute, Kurchatov Square 1, Moscow, 123182 Russia*

**Abstract.** We consider status of the Kr2Det project aimed at sensitive searches for neutrino oscillations in the atmospheric neutrino mass parameter region around $\Delta m^2 \sim 3 \times 10^{-3}$ eV$^2$ and at obtaining new information on the electron neutrino mass structure ($U_{e3}$).

## I. INTRODUCTION

The CHOOZ experiment [1] conclusively showed that oscillations of electron neutrinos do not play the dominant role in the atmospheric neutrino anomaly:

$$sin^2 2\theta_{CHOOZ} \leq 0.1 \quad (\text{at } \Delta m^2 = 3 \times 10^{-3}\ eV^2) \quad (1)$$

In the three active neutrino mixing scheme $sin^2 2\theta_{CHOOZ} = 4\, U^2_{e3}\,(1 - U^2_{e3})$ where $U_{e3}$ is the contribution of the mass-3 eigenstate to the $\nu_e$ flavor state. Thus the CHOOZ result indicates that this contribution is not large:

$$U^2_{e3} \leq 2.4 \times 10^{-2}.$$

The main physical goals of the Kr2Det project [2] are: (i) to look for much smaller mixing angles, find the element $U_{e3}$ or set stronger constraints on its value, (ii) to provide normalization for future long baseline experiments at accelerations and (iii) to reach better understanding of the role $\nu_e$ can play in the atmospheric neutrino anomaly. We mention also that new information on $U_{e3}$ can help to choose between possible solar neutrino oscillation solutions [3].

The main practical goal of the project is to decrease, relative to the CHOOZ, the systematic and statistic errors as much as possible. This can be achieved (i) by using two (far and near) detector scheme of the experiment, which eliminates systematic uncertainties associated with the reactor neutrino flux and spectrum and (ii) by a considerable increase of the number of detected neutrinos.

## 2. The Kr2Det PROJECT

### 2.1. *Detectors*

Two identical liquid scintillation spectrometers stationed at distances $R_1 = 1100$ m (far position) and $R_2 = 150$ m (near position) from the Krasnoyarsk underground (~ 600 mwe) reactor detect (e$^+$, n) pairs, produced in the $\nu_e + p \to n + e^+$ reaction. A miniature version of the KamLAND and BOREXINO [4] three concentric zone detector composition is chosen for the design of the spectrometers (Fig. 1.) The neutrino targets have 46 t of liquid scintillator enclosed in transparent balloons. The target is viewed by ~500 8-inch EMI-9350 PMT's [5] through a ~ 90 cm layer of the isoparaffin of the zone-2. PMT's of the same type were successfully used in the CHOOZ experiment and are used now in the SNO and BOREXINO detectors. A ~120 photoelectron signal is expected for 1 MeV energy deposit in the detector center. The PMT's are mounted on a stainless steel screen, which optically separates the external zone-3 from two central zones. The 75 cm thick zone-3 is filled with mineral oil and serves as active (muons) and passive shielding from the external radioactivity.

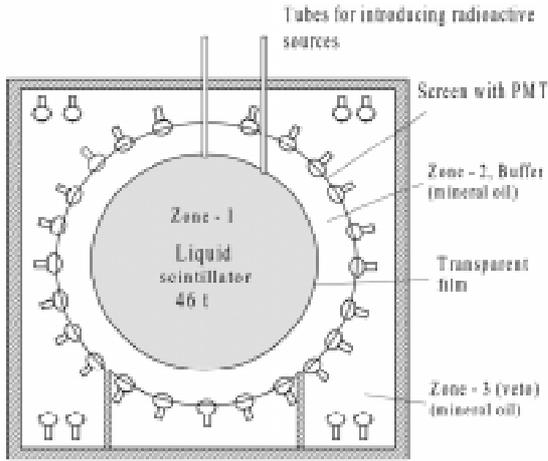

Fig. 1. The detector

## 2.2. *Neutrino rates and backgrounds*

The neutrino event satisfies the following requirements: (i) a time window on the delay between $e^+$ and neutron 2 – 600 μs, (ii) energy window for the neutron candidate 1.7 – 3.0 MeV and for $e^+$ 1.2 – 8.0 MeV, (iii) distance between $e^+$ and neutron less than 100 cm. No pulse shape analysis is planned. Under these assumptions the neutrino detection efficiency ε = 77% was found and neutrino detection rate $N_\nu$= 46 day$^{-1}$ calculated for the detector at 1100 m position.

The expected rate of the correlated background in the Kr2Det project (see Ref. [2]), 0.1 per day per ton of the target mass, is ~ 3 times lower than was measured at CHOOZ, which reasonably agrees with the deeper position (600 vs 300 mwe) of the Kr2Det detectors.

While correlated background originates from cosmic muons, the accidentals come from the natural radioactivity of materials. In Tab. 1 are summarized radioactivity levels of the scintillator and of other detector materials and calculated single rates, which define the rate of accidental coincidences. Backgrounds coming from U and Th contained in the surrounding rock is still under consideration.

Table 1. Single background rates due to natural radioactivity of detector materials.

| Material mass, ton | Isotope | Purity G/g | Background, Hz >1.0 MeV | >1.2 MeV | >1.7 MeV |
|---|---|---|---|---|---|
| Scintillator, 45 ton | $^{238}$U | 10$^{-13}$ | 0.078 | 0.051 | 0.045 |
| | $^{232}$Th | 10$^{-13}$ | 0.033 | 0.022 | 0 017 |
| | $^{40}$K | 10$^{-13}$ | 0.020 | 0.005 | 0.001 |
| | $^{222}$Rn | 1mBq/m3 | 0.050 | 0.040 | 0.040 |
| Σ Scintillator | | | 0.181 | 0.118 | 0.103 |
| Zone-2 oil, 76 ton | $^{238}$U | 10$^{-13}$ | 0.0035 | 0.0019 | 0.0012 |
| | $^{232}$Th | 10$^{-13}$ | 0.0014 | 0.0008 | 0.0007 |
| | $^{40}$K | 10$^{-13}$ | 0.0097 | 0.0033 | 0.0007 |
| | $^{222}$Rn | 10mBq/m3 | 0.033 | 0.018 | 0.0116 |
| Σ zone-2 oil | | | 0.0476 | 0.0240 | 0.0142 |
| PMT, 500 kg | $^{238}$U | 1x10$^{-8}$ | 0.028 | 0.013 | 0.008 |
| | $^{232}$Th | 2x10$^{-8}$ | 0.037 | 0.026 | 0.022 |
| | $^{40}$K | 7x10$^{-9}$ | 0.024 | 0.002 | < 0.001 |
| Σ PMT | | | 0.089 | 0.041 | 0.030 |
| Screen stainless steel 2 ton | $^{238}$U | 5x10$^{-9}$ | 0.056 | 0.27 | 0.016 |
| | $^{232}$Th | 1x10$^{-8}$ | 0.074 | 0.52 | 0.045 |
| | $^{40}$K | 2x10$^{-9}$ | 0.027 | 0.002 | < 0.001 |
| Σ screen | | | 0.157 | 0.081 | 0.061 |
| Zone-3 oil, 218 ton | $^{238}$U | 10$^{-12}$ | | | |
| | $^{232}$Th | 10$^{-12}$ | | < 0.001 | |
| | $^{40}$K | 10$^{-12}$ | | | |
| | $^{222}$Rn | 10mBq/m3 | | | |
| Σ zone-2 oil | | | 0.0027 | 0.0016 | 0.0014 |
| **Total BKG rates** | | | **0.477** | **0.257** | **0.210** |

The calculated accidental background rate is $N_{Acc} \approx 0.1$ day$^{-1}$, which is hundred times smaller than neutrino signal $N_\nu$. We note that BOREXINO and KamLAND detectors aimed to detect solar neutrinos in subMeV range will use scintillators with 3 – 4 of magnitude lower level of radioactive impurities than level sufficient for present project.

Calculated neutrino detection rates and backgrounds are summarized in Tab. 2.

Table 2. Neutrino detection rates and backgrounds

| Parameter | Distance m | Depth mwe | Target mass ton | Neutrino detection rates $N_\nu$, d$^{-1}$ | Backgrounds, d$^{-1}$ correlated | accidental |
|---|---|---|---|---|---|---|
| Far detector | 1100 | 600 | 46 | 46 | 5 | 0.1 |
| Near detector | 150 | 600 | 46 | 2500 | 5 | 0.1 |

*2.3. Data analysis and expected results*

In no oscillation case the ratio of two simultaneously measured positron energy spectra $S_{FAR}/S_{NEAR}$ is energy independent. Small deviations from constant value of this ratio

$$S_{FAR}/S_{NEAR} = C(1 - sin^2 2\theta \cdot sin 2\varphi_F) \times (1 - sin^2 2\theta \cdot sin^2 \varphi_N)^{-1} \qquad (2)$$

are searched for oscillations ($\varphi_{F,N}$ stands for $1.27 \Delta m^2 L_{F,N}/E\nu$ and $L_{F,N}$ are the distances between the reactor and detectors) The results of the analysis do not depend on the exact knowledge of the neutrino flux and their energy spectrum, burn up effects, the numbers of target protons…. However possible relative difference of the detector energy scales should be strictly controlled. This can be done by a combination of different methods briefly mentioned in Ref. [2];

Expected 90% CL oscillation limits are presented in Fig.2. It was assumed that 40000 neutrinos are detected in the far detector and that detector response functions differences are controlled down to 0.5%.

3. CONCLUSIONS

Mass structure of the electron neutrino can sensitively be explored using two detector techniques in underground laboratory.

The project is relatively inexpensive when compared with modern neutrino experiments listed in the *Neutrino Oscillation Industry* www sites. An effective international cooperation would be highly desirable to complete the project and start the experiment.

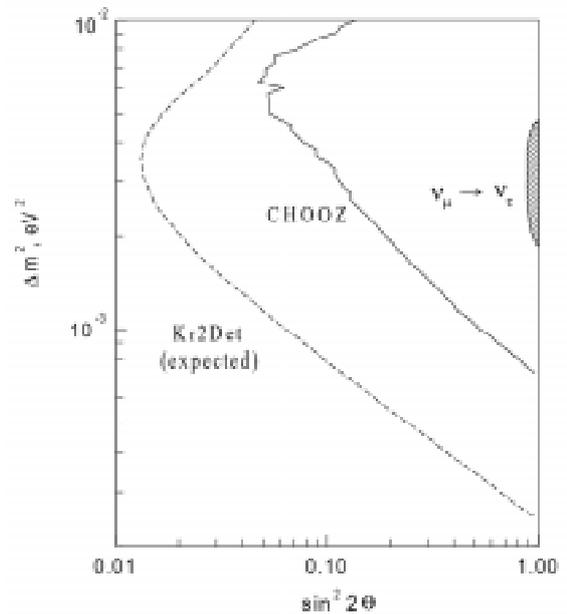

Fig.2. Reactor neutrino 90 % CL disappearance limits. Shaded is the atmospheric $\nu_\mu \to \nu_\tau$ allowed region.

We appreciate valuable discussions with E. Akhmedov, S. Bilenky and A. Smirnov on the physics of neutrino oscillations. Discussions with our colleagues of Kurchatov Institute' neutrino groups are of great help in developing this project. We are thankful to Yu. Kamyshkov and A. Piepke for valuable information on the KamLAND experiment.This study is supported by RFBR grants N 00-02-16035, 00-15-98708.